\documentclass[letter]{aa} 
\usepackage{graphicx}
\usepackage{txfonts}
\def\p1{paper I}
\def\P1{Paper 1}
\begin{document}
   \title{Constraining the rate of GRB visible afterglows with the
CFHTLS Very Wide Survey\thanks{Based on observations obtained with
MegaPrime/MegaCam, a joint project of CFHT and CEA/DAPNIA, at the
Canada-France-Hawaii Telescope (CFHT) which is operated by the
National Research Council (NRC) of Canada, the Institut National des
Sciences de l'Univers of the Centre National de la Recherche
Scientifique (CNRS) of France, and the University of Hawaii}}

   \author{F. Malacrino\inst{1}
       \and
       J-L. Atteia\inst{1}
       \and
       M. Bo\"er\inst{2}
       \and
       A. Klotz\inst{2}$^{,}$\inst{3}
       \and
       C. Veillet\inst{4}
       \and
       J-C. Cuillandre\inst{4} on behalf of the GRB RTAS Collaboration\thanks{Collaboration of researchers who proposed
       the GRB RTAS project (see http://www.cfht.hawaii.edu/$\sim$grb/)}}
\offprints{F. Malacrino, fmalacri@ast.obs-mip.fr}

\institute{Laboratoire d'Astrophysique de Toulouse-Tarbes,
Observatoire Midi-Pyr\'en\'ees
(CNRS-UMR5572/Universit\'e Paul Sabatier Toulouse III), 14 Avenue Edouard Belin, 31400 Toulouse, France\\
          \email{fmalacri@ast.obs-mip.fr}
          \and Observatoire de Haute-Provence, 04870 Saint-Michel l'Observatoire, France
          \and Centre d'Etude Spatiale des Rayonnements, Observatoire Midi-Pyr\'en\'ees
(CNRS/UPS), BP 4346, 31028 Toulouse Cedex 04, France
      \and Canada-France-Hawaii Telescope Corp., Kamuela, HI 96743, USA
         }

   \date{Received ; accepted }


  \abstract
   {}
   {We analyze images of the CFHTLS Very Wide Survey to search for visible
    orphan afterglows from gamma-ray bursts (GRBs).}
   {We have searched 490 square degrees down to magnitude r'=22.5 for 
    visible transients similar to GRB afterglows.
    We translate our observations into
    constraints on the number of GRB visible afterglows in the sky,
    by measuring the detection efficiency of our search with a simulation
    reproducing the characteristics of our observational
    strategy and the properties of on-axis GRB afterglows.
    }
   {We have found only three potential candidates, of which two are most probably variable
    stars, and one presents similarities to an orphan afterglow.
    We constrain the number of visible afterglows to be less than 220 down to
    r'=22.5 in the whole sky at any time. Our observations are marginally 
    consistent with the most optimistic model, which predicts orphan
    afterglows to be about 10 times more frequent than GRBs.}
   {This search has led to the detection of one possible GRB afterglow,
    and provides the strongest constraints on the rate of GRB visible afterglows
    as well as an estimation of the observing
    time required to detect a significant number of GRB afterglows.}

   \keywords{Gamma rays: bursts --
             Methods: data analysis}
   \titlerunning{Constraining GRB afterglows in CFHTLS-VWS}
   \maketitle
%

\section{Introduction}
\label{intro} The prediction of the existence of orphan GRB
afterglows relies on the double assumption that the GRB
prompt emission is beamed and that the afterglow emission
is still bright enough to be detectable when it
starts to radiate outside the GRB beam. This situation, which was described by Rhoads in 1997
soon after the discovery of the first GRB afterglows, makes the detection
of visible afterglows possible even without the GRB trigger.
Such afterglows are usually called 'orphan afterglows'.
The properties and expected number of orphan GRB
afterglows have been discussed by Rhoads
(\cite{rhoads97,rhoads99}), Totani \& Panaitescu (\cite{totani02}),
Nakar, Piran \& Granot (\cite{nakar02}), and Dalal et al.
(\cite{dalal02}). 
Huang et al. (\cite{huang02}) have proposed that failed on-axis
GRBs with Lorentz factors well below 100 could also produce 
orphan afterglows.

\begin{table}[ht]
\caption{Summary of our observations.
The columns give for each filter: the number of images processed, N$_{\rm im}$,
the surface area in square degree, S$_{\rm im}$, the mean accuracy of the astrometry
in arcsecond,
the completeness magnitude, M$_{\rm lim}$, the density of astronomical sources 
per square degree, and the percentage of images that have
been properly processed (see paper I).}
\label{catstat}
\centering
\begin{tabular}{c c c c c c c}
\hline\hline
Filter & N$_{\rm im}$ & S$_{\rm im}$    & Accuracy & M$_{\rm lim}$ & Density        & Success \\
       &          & [$deg^{2}$] & $arcsec$   &   & $deg^{-2}$ &         \\
\hline
g' &  536 &  481 & 0.52 & 23.1 & 31910 & 99.47 \\
r' & 1302 & 1167 & 0.54 & 22.6 & 41370 & 99.30 \\
i' &  589 &  531 & 0.51 & 22.4 & 51075 & 99.83 \\
\hline
\end{tabular}
\end{table}


The detection of orphan afterglows at optical wavelengths would open
a new window in the GRB field. First, it would suppress the existing bias in
GRB studies due to the fact that {\it all} GRBs observed to date
have been detected by their prompt emission at high-energies.
Second, since orphan afterglows are thought to be far more
numerous than on-axis afterglows, we expect to detect many more
nearby afterglows, allowing detailed studies on specific  issues
like the GRB-SN connection.
Third, the detection, or non-detection, of
orphan afterglows will provide constraints on the
beaming angle and energetics of GRBs (see Rhoads \cite{rhoads97},
Totani \& Panaitescu \cite{totani02}, Nakar, Piran \& Granot
\cite{nakar02}).

Given the potential science returns from the detection of visible
orphan afterglows, various searches have been performed, with very
different depth and sky coverage (Becker et al. \cite {becker04},
Rykoff et al. \cite {rykoff05}, Rau et al. \cite {rau06}). These
searches produced no orphan afterglow detection and provided
constraints on GRB beaming which are described in section
\ref{constraints}. We report here the results of a search performed
in images taken for the CFHTLS Very Wide Survey (hereafter CFHTLS-VWS) 
which has a combination of depth and sky coverage
providing an unprecedented sensitivity for orphan afterglow
searches. An extensive description of the survey and of our search
procedure has been given in Malacrino et al. (\cite{malacrino06},
hereafter \p1), we refer the reader to this paper for details. In
this letter we focus on the afterglow candidates found in the
CFHTLS-VWS images (section \ref{results}) and on the constraints that
we derive on the frequency of GRB visible afterglows 
(section \ref{constraints}).


\section{Visible afterglow candidates}
\label{results}

\begin{table*}[ht]
\caption{Properties of our afterglow candidates: Name, Right Ascension, 
Declination, number of images available, date and time of the observations,
magnitude of the source, and a comment
on the most probable nature of the source.} \label{candidates}
\centering
\begin{tabular}{l l l l l l l l l }
\hline\hline
NAME & RA & DEC & N$_{im}$ & Date & Time [SOD] & Filter & Magnitude & Comment \\
 & & & & & & & & \\

\hline

OT 20050629 &19 38 06.81 & - 21 22 31.3 & 4 & 2005-06-29 & 38182 & r' & 21.54 & 
probably a variable star,\\
 & & & & 2005-06-29 & 40527 & r' & 21.55 & only 4 images\\
 & & & & 2005-06-29 & 42924 & r' & 21.37 & \\
 & & & & 2005-07-01 & 42422 & r' & 22.07 & \\

\hline

OT 20050728 &15 57 16.78 & -18 50 58.6 & 11 & 2005-06-10 & 34244 & g' & $>$23.35 & 
afterglow candidate, \\
 & & & & 2005-06-12 & 22204 & g' &  $>$23.19 & detected in 5 out of 11 images \\
 & & & & {\bf 2005-07-28} & {\bf 21849} & {\bf i'} & {\bf 20.08} & \\
 & & & & {\bf 2005-07-28} & {\bf 26054} & {\bf i'} & {\bf 18.66} & \\
 & & & & 2005-07-28 & 30330 & i' & 20.90 & \\
 & & & & {\bf 2005-07-28} & {\bf 30631} & {\bf i'} & {\bf 20.70} & \\
 & & & & {\bf 2005-07-29} & {\bf 22204} & {\bf i'} & {\bf 21.58} & \\

\hline

OT 20060202 &04 54 05.71 & +21 45 17.8 & 4 & 2006-02-02 & 26962 & i' & 21.58 & 
probably a variable star,\\
 & & & & 2006-02-02 & 30226 & i' & 21.74 & only 4 images\\
 & & & & 2006-02-02 & 33356 & i' & 21.91 & \\
 & & & & 2006-02-03 & 25483 & i' & 22.01 & \\
\hline
\end{tabular}
\end{table*}

The present work is based on the analysis of images covering an area
of 490 square degrees.
Their completeness magnitude depends on the filter.
50\% of our images are complete down to g'=23.1, r'=22.6, 
and i'=22.4 (see Table \ref{catstat}). The regions of the sky
that have been observed, and the analysis of the images leading to
the detection and validation of afterglow candidates are described
in detail in \p1. We briefly recall here only the points relevant to our orphan afterglow search.

When a new image is recorded for the CFHTLS-VWS, we construct the
catalog of sources in the image. Usually, three images of the same
field are acquired about 1 hour apart during a single night, and another one on
the following night. Our software compares the sources detected in
these images and searches for
photometrically variable or moving objects, which are checked
by a member of the collaboration less than 24 hours after the acquisition
of the images
(see \p1\ for details). 
A visual examination of these objects led to the rejection of 90\% of them, and to the validation of the remaining
10\% as truly variable objects, which correponds to 0.007\% of the total number of sources.


\begin{figure}[ht]
\begin{center}
\includegraphics[width=1\linewidth,angle=0]{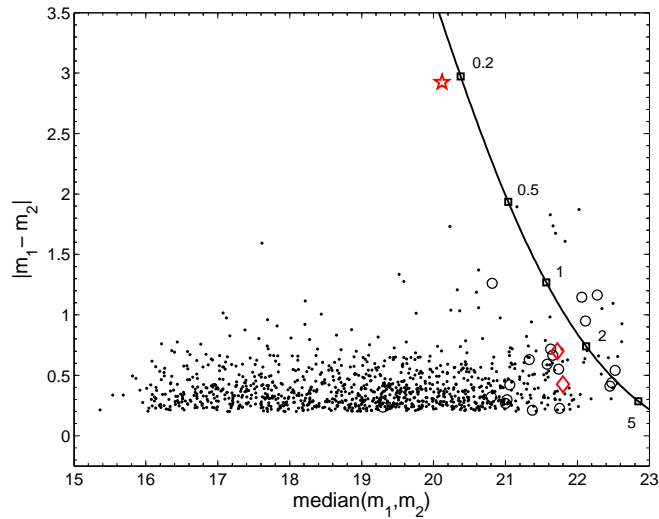}
\end{center}
   \caption{1067 variable objects detected
   by our automatic software and confirmed by visual examination.
   Each point represents a single
object in the inter-night comparison of a pair of images. The x-axis shows the
median magnitude while the y-axis shows the absolute value of the magnitude difference
between the two images.
The open circles and the two red diamonds show the objects that have been
identified as afterglow candidates in the real-time analysis, and discarded (circles),
or not (diamonds) after further analysis (see text).
The solid line shows where a typical afterglow
($\alpha=1.2$, $m_{1}=21$ and $m_{host}=24$) would be placed on this diagram as a
function of its age (in days) at the time of the first observation.
The red star shows OT 20050728, a variable source which has characteristics
similar to GRB afterglows.}
   \label{variables}
\end{figure}

While \p1\ discusses
variable sources found in all comparisons,
{\it we restrict the work in this paper to sources found in inter-night
comparisons}. This is not a strong limitation since all afterglows
should be detected in inter-night comparisons
(see \p1). Another important point is that we have chosen to restrict our
search to objects that vary by 0.2 magnitude or more,
a value that offers a good compromise 
between sensitivity and the number of false detections.
Our analysis has led to the identification of 1067 truly variable objects out of 
more than 18 million, in 549 MegaCAM
fields, which shows that 
few objects in the sky show variations larger than 0.2
magnitude on a timescale of 1 day. The absolute variation in
magnitude of these sources between the two nights as a function of 
their median magnitude is shown in Figure \ref{variables}.

All the objects of figure \ref{variables} have been carefully
examined in order to determine their nature. Most of
them are variable stars which are identified as such because
they are present in archived astronomical 
images\footnote{We use the Aladin sky atlas, Bonnarel et al. (\cite{bonnarel00}).},
or in the USNO-B1.0 catalog (Monet et al. \cite{monet03}),
or in images of the same field acquired
during other observational periods of the CFHTLS-VWS.
This last method is particularly useful for variable objects 
fainter than the limit of the USNO-B1.0 catalog (about $21^{st}$ mag).
This selection assumes that GRB afterglows are truly transient sources.
During this screening process we also eliminate a few slow moving
objects that are referenced in the MPchecker
(http://scully.harvard.edu/$\sim$cgi/CheckMP). These objects can
mimick a GRB afterglow that has 'disappeared' after one day.
Variable sources that are not eliminated by one of these methods 
are classified as 'afterglow candidates'. 

\begin{figure*}[ht]
\begin{center}
\includegraphics[width=0.24\linewidth,angle=270]{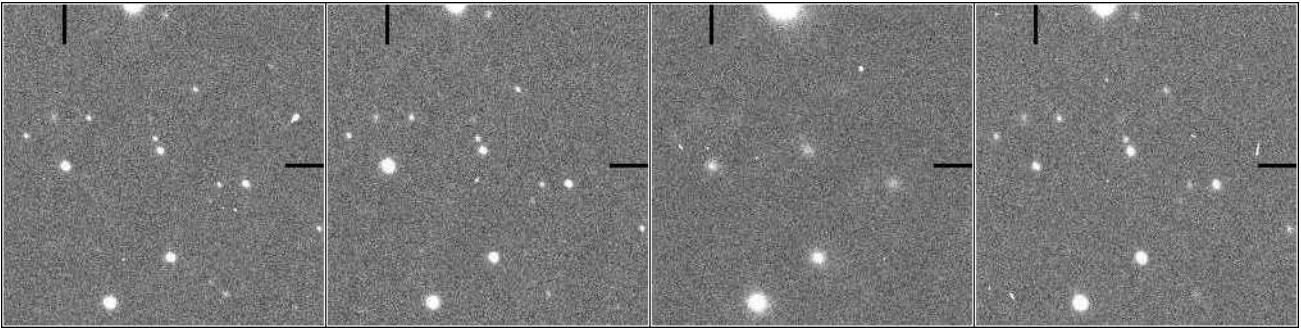}
\end{center}
   \caption{The evolution of OT 20050728 during the nights of 
2005 July 28th and 29th. The 4 images correspond to the 4 lines
in bold in Table \ref{candidates}.}
   \label{ot20050728}
\end{figure*}

During the real-time process, 26 sources were classified
as 'afterglow candidates'. A careful re-analysis of
these sources with the full set of images allowed us to
re-qualify 23 of them as variable stars. These 23 sources
are shown with open circles in Figure \ref{variables}. 
We note that nearly all of them are fainter
than the magnitude limit of USNO-B1.0, emphasizing the role of
this catalog in the rejection of bright variable sources.
At the time of writing this paper we are left with only 3 objects, 
listed in Table \ref{candidates}. Two of them (the diamonds) 
are probably variable stars which we still consider as 'candidates' 
due to the lack of reference images.
The last transient (the red $\star$ in Figure \ref{variables}) is a 
remarkable event which shares some similarities with a visible 
afterglow (the solid line in Figure \ref{variables}
simulates the track of a typical on-axis afterglow).
We have called it OT 20050728, and its evolution is shown in Figure \ref{ot20050728}.
Unfortunately, the available data (see Table \ref{candidates}) are not
sufficient to unambiguously determine its nature. 

One significant source of background in searches for GRB visible afterglows 
is flare stars (Kulkarni \& Rau \cite{kulkarni06}).
We consider that it is unlikely that OT 20050728 is a stellar flare
since its long rise time (greater than 1 hour)
and its 3 magnitude variation in the i' band
are unusual for stellar flares.
We thus consider OT 20050728 as a possible visible afterglow candidate.
In view of the importance
of the characterization of this event, we have requested
additional observations during June and July 2007. We
strongly encourage deep observations of this source.

In the following we discuss the rate of GRB visible afterglows under  
two assumptions; i) our search has led to the detection of
zero afterglow, ii) OT 20050728 is a GRB afterglow.

\section{Constraining GRB beaming}
\label{constraints}

\begin{figure}[ht]
\begin{center}
\includegraphics[width=1\linewidth,angle=0]{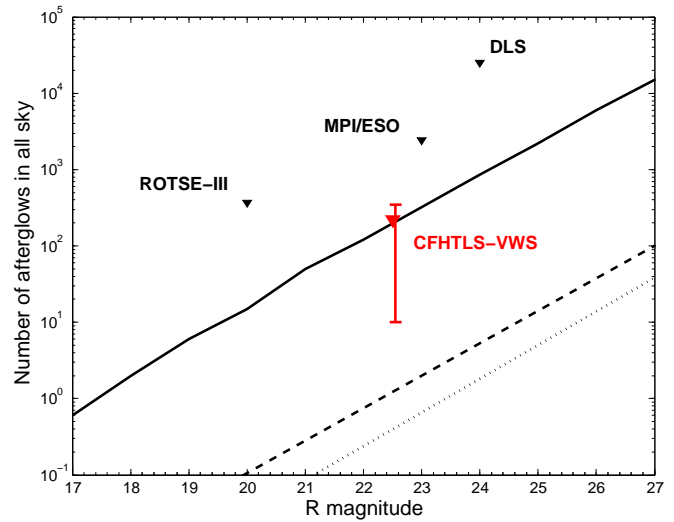}
\end{center}
   \caption{Constraints provided by our observations (VWS),
   compared to the results of ROTSE-III (Rykoff et al. \cite{rykoff05}),
   of the MPI/ESO survey (Rau et al. \cite{rau06}), and
   of the DLS (Becker et al. \cite{becker04}). The upper limit assumes the detection
   of zero afterglow and the error bar the detection of one afterglow in our survey (see
   section \ref{results} for additional explanations). Also shown are the
   theoretical predictions of Totani \& Panaitescu (\cite{totani02}, solid line), 
   of Nakar, Piran \& Granot (\cite{nakar02}, dashed line), and of Zou et al.
   (\cite{zou06}, dotted line).}
   \label{resultatfig}
\end{figure}

In this section we use our observations to constrain 
the number of visible afterglows in the whole sky.
The transformation of this value into a number of {\it detected} afterglows in
a given survey depends on parameters describing the observational
strategy (sky coverage, limiting magnitude, time between consecutive 
observations of a given field...) and
on parameters describing the afterglows (shape 
of the light-curve, magnitude at 1 day, magnitude of the host...).
We have constructed a simple simulation which generates 
random afterglow light-curves and computes the fraction detected in a given observational strategy (see paper I for more details).
One current limitation of our simulation is that it uses {\it observed}
light-curves taken from the GCN Circulars, which are presumably 
generated by on-axis GRBs. This may not be too much of a problem
however because in a deep survey like ours, GRB afterglows are 
detected after several days, when the light-curves of on-axis
and off-axis afterglows are similar.
According to our simulation, and assuming that we have detected no
afterglow, the number of visible afterglows in the sky at a given time down to r'=22.5
is smaller than 220 (90\% confidence). This translates into an upper 
limit of 10100 visible afterglows per year, or less than 13 orphan afterglows
per GRB. If we assume in contrast that we have detected one afterglow, the number of afterglows
in the sky at a given time is 100, with an error bar comprised between 10 and 350 (90\% confidence).
These numbers are reported in Figure \ref{resultatfig} along with
the constraints derived from previous searches and some theoretical estimates.

Figure \ref{resultatfig} shows that our search is about 10 times more
sensitive than previous works. This is due to its unprecedented combination 
of depth and sky coverage which has been permitted by the large area of MegaCAM,
the high throughput of the 3.6 meter CFH Telescope, the organisation
of a survey adequate for visible afterglow searches, 
and by the implementation of dedicated software allowing a quick processing
of the images and an efficient search for variable objects.
If we compare our observations to theoretical predictions,
the assumption that we have detected no afterglow gives an upper limit 
which is  marginally consistent with the (optimistic) prediction 
of Totani \& Panaitescu (\cite{totani02}), and fully consistent
with the more pessimistic predictions of Nakar et al. (\cite{nakar02}),
and Zou et al. (\cite{zou06}). If we assume that our
candidate is a real afterglow we reach a different
conclusion: our data are incompatible with  the predictions
of Nakar et al. (\cite{nakar02}) and Zou et al. (\cite{zou06}) at the
90\% confidence level. More importantly, if OT 20050728 is a true 
GRB afterglow, our work is the first one to provide an estimate 
of the amount of observing time needed to detect a significant 
number of GRB visible afterglows.

While the detection of orphan afterglows was initially proposed
as a way to constrain GRB beaming (Rhoads \cite{rhoads97}), more recent
work has shown that the number of visible afterglows
depends on various factors like the structure of the jet 
(Dalal et al. \cite{dalal02}, Granot et al. \cite{granot02}, 
Nakar et al. \cite{nakar02}, Totani \& Panaitescu \cite{totani02})
and on the existence of orphan afterglows not due to off-axis GRBs 
(Huang et al. \cite{huang02}, Nakar \& Piran \cite{nakar03}). 
It could be possible to constrain the GRB beaming by including 
a model of GRB jets in our simulations, but this is beyond the scope 
of this paper.

Along the duration of the survey our main method to discard 
false alarms in a search for GRB afterglows has been the comparison with images taken months or years 
earlier or later. In doing so we rely on the truly transient nature of GRB 
afterglows, which is one of their most unique properties.
This is illustrated in Figure \ref{variables}, which shows 
that, among 22 million objects, we have found one thousand 
sources varying by more than 0.2 magnitude in one day,  
and that only one of these sources was truly transient. 
We believe that future efficient searches should be made in regions 
of the sky covering several tens to a few hundred square degrees, 
already observed down to m $\sim$ 24-25, and which are outside the
Ecliptic and Galactic planes. This is typically the case of weak shear 
surveys, like the CFHTLS Wide Synoptic Survey which covers a total of 
175 square degrees in 5 filters (u*,g',r',i',z'), down to i'=24.5.
The existence of such catalogs represents a remarkable opportunity
for future searches of GRB visible afterglows. Moreover, the combination
of optical and radio observations (Levinson et al. \cite{levinson02},
Gal-Yam et al. \cite{galyam06}) may be the best way to measure GRB beaming in the near future.




\begin{acknowledgements}
We thank everyone at CFHT for their continuous support, 
especially Kanoa Withington. We thank the Observatoire
Midi-Pyr\'en\'ees for funding the Real Time Analysis System.
\end{acknowledgements}

\end{document}